\input{aipcheck}

\documentclass[
    ,final            % use final for the camera ready runs
  ]
  {aipproc}

\layoutstyle{8x11single}

\usepackage[koi8-u,engukr]{statphys09}

\usepackage{amssymb}
\usepackage{bm}
\usepackage{mathtext}
\begin{document}

\title{Phase diagrams of the Bose-Hubbard model at finite temperature}


\author{I.V. Stasyuk}
{
  address={ Institute for Condensed Matter Physics,1 Svientsitskii Str., 79011 Lviv, Ukraine  }
}

\author{T.S. Mysakovych}
{
%  address={Institute for Condensed Matter Physics,1 Svientsitskii Str., 29011 Lviv, Ukraine  }
}

%\author{<author3>}{
%  address={<common address for author2 and author3>}
%  ,altaddress={<author1 address>} % additional visiting address
%}

\begin{abstract}

The phase transitions in the  Bose-Hubbard model are investigated.
     A single-particle Green's function is calculated
   in the
   random phase approximation and the formalism of the Hubbard operators is used.
         The regions of existence of the superfluid and Mott insulator  phases are
          established and the 
$(\mu,t)$ (the chemical potential -- transfer parameter)  
phase diagrams 
      are built.
           The influence
   of temperature change on this transition is analyzed and the phase
    diagram in the $(T,\mu)$ plane is constructed.
    The role of thermal activation of the ion hopping
     is investigated by taking into account the temperature dependence of the
    transfer parameter.
    The reconstruction of the
      Mott-insulator lobes due to this effect is analyzed.

\end{abstract}

\maketitle

%%%%%%%%%%%%%%%%%%%%%%%%%%%%%%%%%%%%%%%%%%%%
%% MAINMATTER
%%%%%%%%%%%%%%%%%%%%%%%%%%%%%%%%%%%%%%%%%%%%

\section{Introduction}

 The Bose-Hubbard model (BHM) has been intensively investigated in the last 15 years. This model 
 is of great interest  due to the experimental realization of Bose-Einstein condensation of ultra-cold atoms in optical lattices (see, for example,  \cite{greiner}). 
 Various theoretical methods were used to study the model: mean-field theory \cite{krish}, random phase approximation \cite{konabe,matsumoto}, strong coupling approach \cite{freericks}, quantum Monte-Carlo method \cite{monte},   bosonic version of dynamical mean field theory  \cite{bychzuk}.
 The Hamiltonian of the model includes two terms. One is connected with the on-site energy $U$, which describes repulsion of bosons at a lattice site; another one 
 describes the 
 nearest-neighbor hopping with tunneling parameter $t$:
 \begin{eqnarray}\label{ham}
 H&=&-\sum_{ij} t_{ij} b^+_i b_j 
+ \frac{U}{2}\sum_{i}n_i(n_i-1)
 - \mu  \sum_{i}  n_{i},
 \end{eqnarray} 
 $\mu$ is the chemical potential. 
 The competition between the kinetic and on-site repulsion  terms defines the equilibrium state of the system. 
 The existence of the superfluid and Mott-insulator phases is a characteristic feature of this model.
When the kinetic energy dominates ($t/U\gg 1$) the ground state of the system  is superfluid (SF). In the opposite case the ground state is a Mott insulator (MI).

 In the limit $U\rightarrow \infty$ this model reduces to the hard-core boson model.
 Models of a lattice gas type with Pauli statistics of particles are often used  for the description of ionic conductors and   calculation of their conductivity
starting from the works of Mahan \cite{mahan} and others \cite{tomoyose,dulepa}.
 Recently  the one-particle spectrum has been investigated in the one-dimensional limit \cite{dulepa}.
 As noted above, the Bose-Hubbard model can be directly applied to the description of optical lattices. Besides, the model of Bose-Hubbard type can be also usefull not only for investigation of ionic conductivity in crystalline ionic conductors, but it can also be  applied to describe the intercalation in crystals  \cite{last2008} and kinetics of ionic adsorption on the crystal surfaces \cite{reilly}.

 In this work we consider the case of the finite value of the on-site interaction $U$ and 
study  
 phase transitions in the BHM at finite temperatures. 
 The most  previous investigations of the Bose-Hubbard model were 
restricted to the case of zero temperature. Recently only the low-temperature case ($T\ll U$) has been investigated  \cite{konabe,pelster}.
A special attention will be paid here to the investigation of the influence of  thermal activation of the ion hopping  on the shape of phase diagrams (we take into account  the temperature dependence of the transfer parameter).
 The effect of thermal activation can be important 
 in the case of ionic conductors. It
was  not  studied in the framework of the BHM; such an investigation has been performed up to now only in the hard-core boson limit  \cite{dulepa}.

%In our previous work
% \cite{last2008} we studied the phase transitions in the  pseudospin-electron model 
% in the case when ions obey Pauli spin-1/2 commutation rules (hard-core boson case) and took into account %the  ion-electron interaction and possibility of ion hopping between neighbour sites. In the present
% investigation  we suppose that more than one ion can be situated at a site and it is regulated by the value %of the on-site interaction $U$.
% In this work we do not consider electron subsystem, this is the task for future investigations.

\section{The random phase approximation}

 We introduce the on-site basis $|n>_i$ ($n$ is a number of bosons on a site $i$) and use the Hubbard operators 
 $X^{nm}_i=|n>_i<m|_i$.  
 The Bose-operators of creation and annihilation can be expressed in terms of the X-operators:
 \begin{eqnarray}\label{x_oper}
 &&b_i=\sum_{n} \sqrt{n+1} X_i^{n,n+1}, \ \ 
 n_i=\sum_n n X^{nn}_i.
 \end{eqnarray} 
 Then the Hamiltonian (\ref{ham}) can be written as follows:
 \begin{eqnarray}\label{ham_d}
 H&=& H_0+H_1, \nonumber\\
H_0&=&\sum_{i n} \lambda_{n } X^{nn}_i;  \   \  \ 
 \lambda_{n }=\frac{U}{2}n(n-1)-\mu n, \nonumber\\
 H_1&=&-\sum_{ij}t_{ij}b^+_i b_j.
 \end{eqnarray}

 To find the Green's function $\langle\langle b|b^+ \rangle\rangle$  we employ the equation of motion technique:
 \begin{eqnarray}\label{equation}
 \omega\langle\langle X^{m,m+1}_l|X^{r+1,r}_p\rangle\rangle&=&
 \frac{1}{2\pi} \delta_{lp} \delta_{mr} \langle X^{mr}_l-X^{r+1,m+1}_l\rangle +
 (\lambda_{m+1}-\lambda_m)\langle\langle X^{m,m+1}_l|X^{r+1,r}_l\rangle\rangle\nonumber\\
& -&\sum_{ij} t_{ij} \sqrt{m+1}\delta_{li} \langle\langle(X^{mm}_l-X^{m+1,m+1}_l)b_j|X^{r+1,r}_p
\rangle\rangle\nonumber\\
&-&\sum_{ij}t_{ij}\delta_{lj}\langle\langle b^+_i(\sqrt{m+2}X^{m,m+2}_l-
 \sqrt{m}X^{m-1,m+1}_l) |X^{r+1,r}_p \rangle\rangle.
 \end{eqnarray} 

 We use the random phase approximation (which is an analogy to the Hubbard-I approximation for the case of the fermionic Hubbard model) and perform the following decoupling:
 \begin{eqnarray}\label{rpa}
 \langle\langle 
(X^{m,m}_l-X^{m+1,m+1}_l)b_j|X^{r+1,r}_p\rangle\rangle
 &\approx& \langle X^{mm}-X^{m+1,m+1} \rangle 
 \langle\langle b_j |X^{r+1,r}_p \rangle\rangle  \nonumber\\
 \langle\langle b^+_i X^{m,m+2}_l|X^{r+1,r}_p \rangle\rangle &\approx& 0.
 \end{eqnarray} 
  It should be noted  that here we consider the Mott-insulator phase, where $\langle b_i\rangle =0$. 
 If we want to investigate the superfluid phase we should take into account that in this phase 
$\langle b_i\rangle \neq 0$ (this case is considered below).

 As a result,  the following expression for the Green's function is obtained:
\begin{eqnarray}\label{gr1}
 &&\langle\langle X^{m,m+1}_l|X^{r+1,r}_p\rangle\rangle=
 \frac{1}{2\pi} \delta_{lp} \delta_{mr}
 \frac{Q_m}{\omega-\Delta_m}-
 \frac{Q_m}{\omega-\Delta_m}\sqrt{m+1}
\sum_j t_{lj} \langle\langle b_j|X^{r+1,r}_p  \rangle\rangle, \\
 &&\Delta_m=\lambda_{m+1}-\lambda_m, \ 
 \langle X^{mm}-X^{m+1,m+1}\rangle=Q_m.
 \end{eqnarray} 
 We pass to $\bm{k}$-representation and the Green's function 
 $G_{\bm k} (\omega)\equiv \langle\langle b|b^+   \rangle\rangle_{\bm k\omega}$
 can be written as
 \begin{eqnarray}\label{g_k}
G_{\bm k}(\omega)&=&   
 \frac{1}{2\pi} \frac{g^0}{1+g^0 t_{\bm k}}\nonumber\\
 g^0(\omega)&=&\sum_{m}\frac{Q_m (m+1)} {\omega-\Delta_m}.
 \end{eqnarray} 
 We should note that  in the independent subband approximation  (this approximation is valid in the case 
$U\gg t$ and in the regions of the integer value of $\mu/U$) the Green's function takes the form
 \begin{eqnarray}\label{indep}
 G_{\bm k}&=&   
 \frac{1}{2\pi}\sum_m \frac{(m+1) Q_m}
 {\omega-\Delta_m+t_{\bm k} (m+1)Q_m},
 \end{eqnarray} 
 and 
  the subband energy  is equal to $\lambda_m=\Delta_m-(m+1)Q_m t_{\bm k}$. 

 In order to study  the phase with the Bose-Einstein condensate we can also apply decoupling of the
 (\ref{rpa}) type, which in this case is more complicated:
 \begin{eqnarray}\label{rpa2}
 \langle\langle 
(X^{m,m}_l-X^{m+1,m+1}_l)b_j|X^{r+1,r}_p\rangle\rangle
 \approx Q_m \langle\langle b_j |X^{r+1,r}_p \rangle\rangle+
 \langle b_j \rangle
\langle\langle (X^{mm}_l-X^{m+1,m+1}_l)|X^{r+1,r}_p \rangle\rangle.
 \end{eqnarray} 
The function 
 $\langle\langle 
(X^{m,m}_l-X^{m+1,m+1}_l)b_j|X^{r+1,r}_p\rangle\rangle$ satisfies the following equation:
 \begin{eqnarray}\label{equation2}
 &&\omega\langle\langle X^{m,m}_l-X^{m+1,m+1}_l|X^{r+1,r}_p\rangle\rangle=
 \frac{1}{2\pi} \delta_{lp} (\langle X^{m,m-1}\rangle\delta_{m,r+1}
+\langle X^{m+2,m+1}\rangle \delta_{m+1,r}\nonumber\\&&
-
2\langle X^{m+1,m}\rangle\delta_{mr})-
\sum_j t_{lj} {\Big(}\sqrt{m}\langle\langle X^{m,m-1}_l b_j- b^+_j X^{m-1,m}_l|X^{r+1,r}_p\rangle\rangle
\nonumber\\&& +\sqrt{m+2} \langle\langle X^{m+2,m+1}_l b_j- b^+_j X^{m+1,m+2}_l|X^{r+1,r}_p\rangle\rangle
\nonumber\\&&
-2\sqrt{m+1} \langle\langle X^{m+1,m}_l b_j- b^+_j X^{m,m+1}_l|X^{r+1,r}_p\rangle\rangle{\Big )}.
 \end{eqnarray} 
Using decoupling (\ref{rpa2}) in the independent subband approximation (which is equivalent to  taking into consideration the transition between levels $|m\rangle$ and $|m+1\rangle$ only)
 the equation takes the more simple form
\begin{eqnarray}\label{1s}
 &&\omega\langle\langle X^{m,m}_l-X^{m+1,m+1}_l|X^{r+1,r}_p\rangle\rangle=
 -\frac{1}{\pi} \delta_{lp}\delta_{mr} (\langle X^{m+1,m}\rangle
 \nonumber\\
&& 
+2(m+1)
\sum_j t_{lj} (\langle X^{m,m+1}_j \rangle\langle\langle X^{m+1,m}_l |X^{r+1,r}_p\rangle\rangle
-
\langle X^{m,m+1}_l \rangle\langle\langle X^{m+1,m}_j |X^{r+1,r}_p\rangle\rangle)
\nonumber\\
&&-
2(m+1)
\sum_j t_{lj} (\langle X^{m+1,m}_j \rangle\langle\langle X^{m,m+1}_l |X^{r+1,r}_p\rangle\rangle
-
\langle X^{m+1,m}_l \rangle\langle\langle X^{m,m+1}_j |X^{r+1,r}_p\rangle\rangle).
\end{eqnarray}
 
 In a similar way, we can obtain equations for the Green's functions 
 $\langle\langle X^{m+1,m}_l |X^{r+1,r}_p\rangle\rangle$ and 
$\langle\langle X^{m,m+1}_l |X^{r+1,r}_p\rangle\rangle$:
\begin{eqnarray}\label{2s}
 &&(\omega+\Delta_m)\langle\langle X^{m+1,m}_l |X^{r+1,r}_p\rangle\rangle=
(m+1)\sum_j t_{lj}Q_m \langle\langle X^{m+1,m}_j |X^{r+1,r}_p\rangle\rangle\nonumber\\&&+
(m+1)\sum_j t_{lj}
\langle X^{m+1,m}_j \rangle
 \langle\langle X^{m,m}_l - X^{m+1,m+1}_l |X^{r+1,r}_p\rangle\rangle.
\end{eqnarray}
\begin{eqnarray}\label{3s}
 &&(\omega-\Delta_m)\langle\langle X^{m,m+1}_l |X^{r+1,r}_p\rangle\rangle=
\frac{1}{2\pi} \delta_{lp} \delta_{mr} Q_m 
 - 
(m+1)\sum_j t_{lj}Q_m \langle\langle X^{m,m+1}_j |X^{r+1,r}_p\rangle\rangle
\nonumber\\&&
-
(m+1)\sum_j t_{lj}
\langle X^{m,m+1}_j \rangle
 \langle\langle (X^{m,m}_l - X^{m+1,m+1}_l) |X^{r+1,r}_p\rangle\rangle.
\end{eqnarray}

The levels $|m>,|m+1>$ form two-state subspace where we can introduce pseudospin formalism:
\begin{eqnarray}
X^{m,m+1}_l=S^-_{l,m};  \  X^{m+1,m}_l=S^+_{l,m}; 
 \   
 \frac{1}{2}\langle X^{m+1,m+1}_l - X^{mm}_l\rangle=S^z_{l,m}.
\end{eqnarray}
 The mean value $Q_m=-2\langle S^z_m\rangle$ is connected with the longitudinal component of the pseudospin, while 
 the order parameter $\langle b_j \rangle  $ in the condensate phase is proportional to the transverse component
 $\langle b_j\rangle \approx \sqrt{m+1} \langle S^-_{m} \rangle=
\sqrt{m+1}\langle S^x_{m} \rangle.$
 
 The set of equations (\ref{1s}), (\ref{2s}), and (\ref{3s}) can be rewritten using pseudospin Green's functions.
 The equations are analogous to those obtained in the RPA for the XXZ model. Their solution can be written in the form given in 
 \cite{dulepa2}
\begin{eqnarray}
\langle\langle S^+_m|S^-_m\rangle\rangle&=&
 \frac{1}{2\pi} \langle \sigma^z_m\rangle
 \frac{E_m(\cos^2\theta_m+1)+2\omega\cos\theta_m-4\langle\sigma^z_m \rangle J_{q,m}\cos^2\theta_m}
{\omega^2-(E_m-2 \langle \sigma^z_m \rangle J_{q,m})(E_m-2 \langle \sigma^z_m \rangle J_{q,m})
\cos^2\theta_m }.
\end{eqnarray}
 In the mean field approximation the angle $\theta_m$ and parameter $\langle\sigma^z_m \rangle$ are defined by the following equations:
\begin{eqnarray}
&&\Delta_m \sin \theta_m-2 (m+1)t_0 \langle S^x_m \rangle \cos\theta_m=0\\
&&\langle \sigma^z_m \rangle=\frac{1}{2}\tanh \frac{\beta E_m}{2}; \  \  \   E_m=\sqrt{\Delta^2_m+4(m+1)^2t^2_0 \langle S^x_m \rangle^2}; \nonumber\\
 &&t_0\equiv t_{q=0};  \  \  
 J_{q,m}=(m+1)t_q;  \   \ 
 \langle S^x_m \rangle=-\langle \sigma^z_m \rangle\sin\theta_m.
\end{eqnarray}
 The solution $\sin\theta_m=0$ ($\langle S^x_m \rangle=0$) corresponds to the normal (nonsuperfluid)
 phase; in the phase with condensate $\cos \theta_m = -\frac{\Delta_m}{2(m+1)t_0
 \langle \sigma^z_m\rangle  }$ and $\langle \sigma^z_m\rangle$ is given by the solution of the following equation
\begin{eqnarray}
&&\langle\sigma^z_m \rangle=\frac{1}{2}\tanh( \beta(m+1)t_0 \langle\sigma^z_m \rangle).
\end{eqnarray}
 
 The transition to the phase with condensate takes place at temperature
\begin{eqnarray}\label{T_c}
&&T^c_m=
 \Delta_m \Big(\ln\frac{1+\frac{\Delta_m}{J_{0,m}}}
{1-\frac{\Delta_m}{J_{0,m}}}\Big)^{-1}.
\end{eqnarray}
This temperature reaches its maximum value $T^c_{max}=J_{0,m}/2$ at
 $\Delta_m=mU-\mu=0$.

 The case of  pair of the states $|0>,|1>$ corresponds to the hard core boson model; the peculiarities of the spectrum and  density of states  were investigated in \cite{dulepa2}.  The thermodynamics of the model  was also studied in \cite{dulepa2} in the case of the thermal activated hopping.

\section{Results}

 The phase transition from the Mott-insulator to superfluid phase is 
characterized by
 divergence of the Green's function $G_{{\bm k}=0}(\omega=0)\rightarrow\infty$.
 This condition can be obtained using the following reasoning.
 Let us find the expression for the commutator $[X^{m,m+1}_l,H]$ in the Hubbard-I type approximation:
 \begin{eqnarray}\label{divergence}
 [X^{m,m+1}_l,H]\approx \Delta_m X^{m,m+1}_l-
 \sum_j t_{lj}\sqrt{m+1}\langle X^{mm}-X^{m+1,m+1}\rangle b_j.
  \end{eqnarray} 
The mean value of this commutator should be equal to zero (because the average $\langle X^{m,m+1}
 \rangle$ does not depend on time in a stationary state). We get after decoulping:
 \begin{eqnarray}\label{condition}
 \langle b_{\bm k}\rangle[1-t_{\bm k} \sum_m \frac{m+1}{\Delta_m} Q_m]=0.
  \end{eqnarray} 
Thus, we can see that  the
 nonzero solution 
 $\langle b_{\bm k} \rangle \neq 0 $ 
 appears in
the point of divergence of the Green's function $G_{{\bm k}}(\omega=0)\rightarrow\infty$ (\ref{g_k}).
The condition (\ref{condition}) in the case $t>0$ (considered in this paper) is fulfilled at the center of the Brillouin zone when ${\bm k}=0$ (an uniform phase case) and  $G_{{\bm k}=0}(\omega=0)\rightarrow\infty$  
  in the phase transition point. In the case $t<0$ this condition is satisfied at the Brillouin zone edge when  ${\bm k}=(\pi,\pi,...) $ (a modulation phase case).

In our numerical calculations  we used the diagonal Hamiltonian $H_0$ to calculate the averages $\langle X^{mm} \rangle$. In this case the condition (\ref{condition}) coincides with that obtained in \cite{pelster} in the framework of the diagrammatic hopping expansion at the calculation of temperature Green's functions (the first order in the self-energy in the notations used in \cite{pelster}).

  Let us consider the influence of thermal activation. We rewrite the parameter of the ion hopping in the following form:  $t=t_0 \exp(-\beta\Delta)$.  Such an approximation was used in \cite{mahan,dulepa} to take into account the presence (in ionic conductors) of a static barrier over which a particle must hop. 
 It can be obtained also  as a result of renormalization due to the interaction with phonons.

\begin{figure}[!htb]
%\centerline{
\includegraphics[width=7cm]{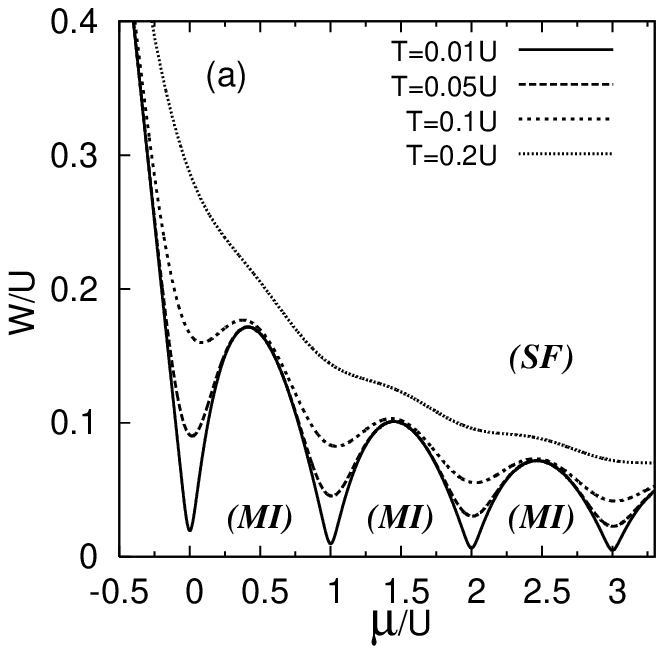}   
\includegraphics[width=7cm]{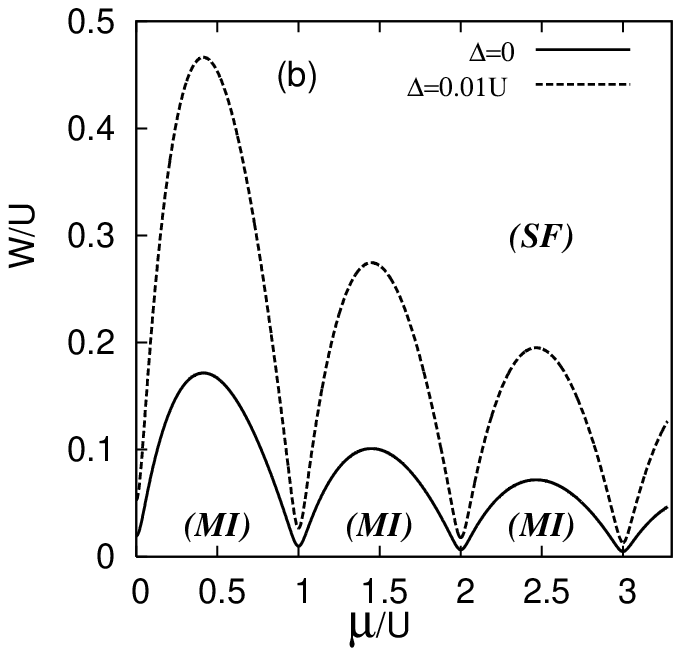}
%}
%\centerline{
%
%\hspace*{-9cm}
%(a)  
%\hspace*{5cm}
% (b)
%}
\caption{  The $(W,\mu)$ phase diagram  at $\Delta=0$ (a) and $\Delta\neq0$ (b). 
 }
\label{11}
\end{figure}

 In Fig.~\ref{11} the ($W,\mu$) phase diagram is shown ($W$ is a half width of the initial energy band, $-W<t<W$). The system can pass from the MI to SF phase at the change of the chemical potential as well as at the change of the transfer parameter.
The influence of thermal activation on phase transitions is illustrated in Fig.\ref{11}b.
 Thermal activation leads to expanding  of the region of the Mott-insulator phase.

\begin{figure}[!htb]
%\centerline{
\includegraphics[width=7cm]{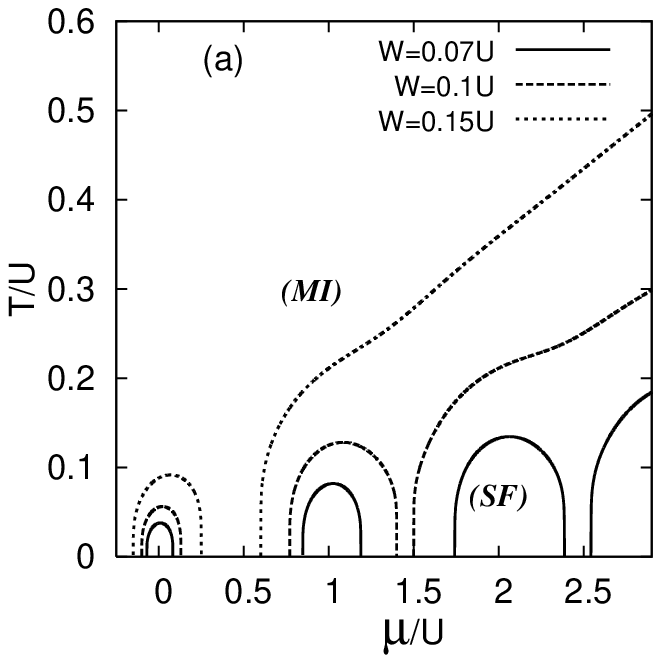}  
\includegraphics[width=7cm]{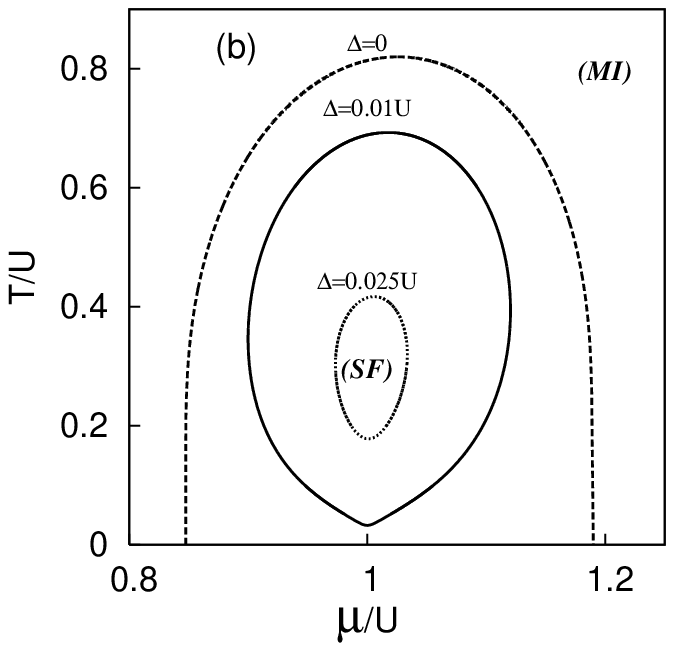}
%}
%\centerline{
%(a)  \hspace{5cm}  (b)  
%}
\caption{ 
   The $(T,\mu)$ phase diagram at $\Delta=0$ (a) and $\Delta\neq0$, $W=0.07U$ (b).
 }
\label{2}
\end{figure}

 The $(T,\mu)$ phase diagrams are given in Fig.\ref{2}.  The influence of thermal activation is 
illustrated   in Fig.\ref{2}b in the case of the one SF lobe ($\mu/U\approx 1$).  At the increase of the activation energy this region narrows and
then the superfluid phase disappears
 (there is a critical value $\Delta_{cr}$). 
It should be noted that at the fixed value of the chemical potential there are two  critical temperatures $T_{c_1}$ and $T_{c_2}$ at which the phase transitions between the MI  and superfluid phase occur (it takes place at $\Delta\neq 0$). One can think that the lower  critical temperature $T_{c_2}$ can be connected with the temperature of the transition to the superionic state in the case of superionic crystals. 
 The simple estimate, based on the typical values of $t$ and $\Delta$ parameters for proton conductors 
 ($\Delta\sim 0.1 ...0.4$ eV; $t\sim 0.2 ... 0.5$ eV) shows that $T_{c_2}$ can be of the order of several hundreds of Kelvins, that corresponds to temperatures observed in the known family M$_3$H(XO$_4$)$_2$  (M=NH$_4$, Rb, Cs; X=S,Se) of hydrogen-bonded superionic crystals \cite{pavl}.

\begin{figure}[!htb]
%\centerline{
\includegraphics[width=7cm]{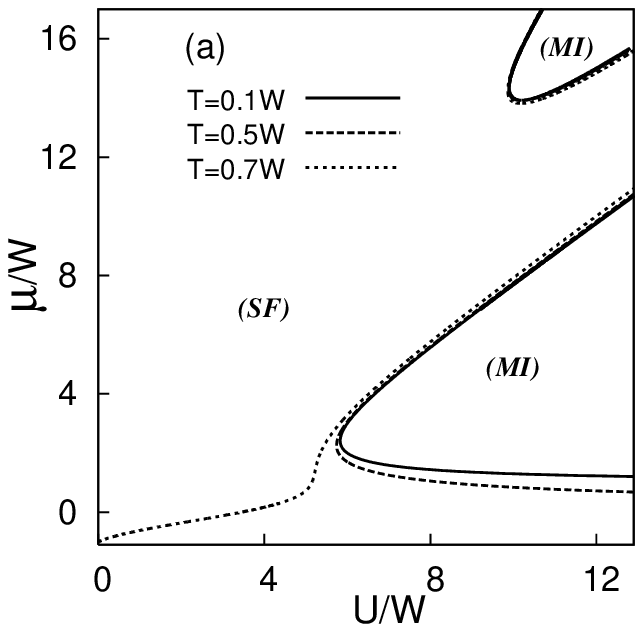} 
\includegraphics[width=7cm]{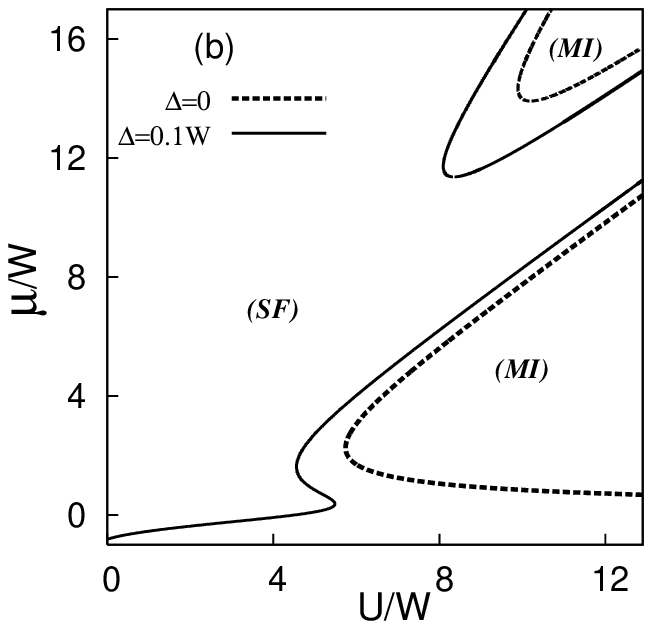}
%}
%\centerline{
%(a)  \hspace{5cm}  (b) 
% }
\caption{ 
 The $(U,\mu)$ phase diagram at $\Delta=0$ (a) and $\Delta\neq0$, $T=0.5W$
  (b).
}
\label{3}
\end{figure}

 From  the $(U,\mu)$ phase diagrams  shown in Fig.\ref{3} 
 we can see that the influence of thermal activation on the phase diargams is similar to the influence of temperature change 
in the case $\Delta=0$.

\begin{figure}[!htb]
%\centerline{
\includegraphics[width=7cm]{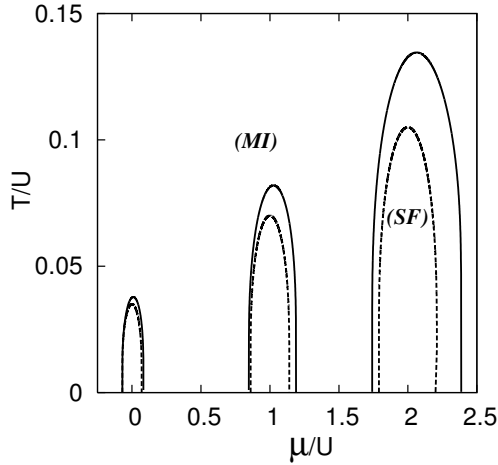} 
%}
%\centerline{(a)  \hspace{5cm} (b)}
\caption{Dashed line denotes the phase transition line obtained in the independent subband approximation
 using (\ref{T_c}). Solid line denotes the phase transition line obtained using 
(\ref{condition}). $W=0.07U, \Delta=0$.
 }
\label{1}
\end{figure}

 In Fig.\ref{1} the $(T,\mu)$ phase diagram calculated in the independent subband approximation 
 using the equation (\ref{T_c}) (where $J_{0,m}=(m+1)t_0 \exp(-\beta^m_c \Delta )$) is shown (dashed line). Solid line is the line shown in Fig.\ref{2}a in the case $W=0.07U$. 
We can see that the independent subband approximation 
 is applicable
 at small values of $m$. In this region the expression (\ref{T_c}) describes adequately the nearly linear growth of $T^m_c$ with the increase of the number $m$.

 \section{Conclusions}

 The phase transitions in the Bose-Hubbard model at finite temperature are investigated.
   Models of such type can  be applied for the description of optical lattices, 
       ionic conductivity,
         intercalation in crystals, and
         kinetics of ionic adsorption on the
         crystal surfaces.

     A single-particle Green's function is calculated
   in the 
   random phase approximation (which is an analogy to the Hubbard-I approximation in the
    case of the fermionic Hubbard model) and the formalism of the Hubbard operators is used.  
 This Green's function is obtained for the  Mott-insulator and superfluid phases (in the last case we use the independent subband approximation, which is appropriate for the case of the noninteger (intermediate) mean number of the particle concentration per lattice site). 
%    Single-particle excitations in the Mott insulator phase are studied
%         (the existence of the energy gap at $\vec{k}=0$ which vanishes
%          at ctitical point is a feature of this phase, reflecting the
%           localized character of atoms) for the case when the interaction
%           between three nearest bands is taken into account.

         The regions of existence of the superfluid and Mott insulator  phases are
          established and the
  phase diagrams in the plane
   $(\mu,t)$ (the chemical potential -- transfer parameter)
 are built.
           The influence
   of temperature change on this transition is analysed and the phase
    diagram in the $(T,\mu)$ plane is constructed.
    The role of thermal activation of the ion hopping 
     is investigated by taking into account the temperature dependence of the 
transfer parameter ($t=t_0 \exp(-\beta\Delta)$).
    It is shown that thermal activation leads to the reconstruction
     of the 
      Mott-insulator lobes (in particular, to the narrowing of the SF phase region). It is revealed that due to  thermal activation of the ion hopping, 
there are two critical  temperatures
 at which the MI-SF phase transition takes place (at the fixed value of the chemical potential).
 A simple estimation shows that the lower temperature can be  quite high  in proton conductors and
 can correspond to the temperature of the superionic phase transition in such compounds.

%% The following lines show an example how to produce a bibliography
%% without the help of the BibTeX program.
%%%%%%%%%%%%%%%%%%%%%%%%%%%%%%%%%%%%%%%%%%%

%%%%%%%%%%%%%%%%%%%%%%%%%%%%%%%%%%%%%%%%%%%%%%%%
%%
%% The code below assumes that BibTeX is used. This could be used instead
%% of the above. If the bibliography is
%% produced with BibTeX uncomment out the following lines and see the

%\bibliographystyle{aipproc}  % if natbib is available
%%   or

%\bibliography{sample}

\begin{thebibliography}{18}







\bibitem{greiner}
  M.~Greiner, O.~Mandel, T.~Esslinger, T.~W. H{\"a}nsch, I.~Bloch, \emph{ Nature} \textbf {415}, 39--44 (2002).



\bibitem{krish}
  K.~Sheshadri, H.~R. Krishnamurthy,   R.~Pandit, T.~V. Ramakrishnan,
  \emph {Europhys. Lett.} \textbf {22}, 257--263 (1993).



\bibitem{konabe}
  S.~Konabe, T.~Nikuni, M.~Nakamura, \emph {Phys. Rev. A} \textbf  {73 }, 033621--12 (2006).

\bibitem{matsumoto}
 Y.~Ohashi, M.~Kitauri,  H.~Matsumoto, \emph {Phys. Rev. A} \textbf  {73}, 033617--5 (2006).

\bibitem{freericks}
  J.~K. Freericks, H.~Monien, \emph {Europhys. Lett.} \textbf {26}, 545--550  (1994).


\bibitem{monte}
  B.~Capogrosso-Sansone, S.~G. S{\" o}yler, N.~Prokof'ev,
B.~Svistunov, \emph {Phys. Rev. A} \textbf  {77}, 015602--4 (2008).




\bibitem{bychzuk}
     K.~Byczuk, D.~Vollhardt, \emph {Phys. Rev.  B} \textbf { 77}, 235106--14   (2008).





 \bibitem{mahan}
  G.~D. Mahan, \emph {Phys. Rev. B} \textbf {14}, 780--793 (1976).

 \bibitem{tomoyose}
 T.~Tomoyose, \emph {J. Phys. Soc. Jpn.} \textbf {66}, 2383--2385 (1997). 
 


\bibitem{dulepa}
  I.~V. Stasyuk, I.~R. Dulepa, \emph {Condens. Matter Phys.} \textbf {10}, 259--268 (2007).



\bibitem{last2008}
 T.~S. Mysakovych, V.~O. Krasnov, I.~V. Stasyuk, \emph {Condens. Matter Phys.} \textbf {11}, 663--667 (2008).


\bibitem{reilly}
 P.~D. Reilly, R.~A. Harris, K.~B. Whaley, \emph {J. Chem.  Phys.} \textbf {95}, 8599--8615 (1991).






\bibitem{pelster}
 M.~Ohliger, A.~Pelster, \emph {cond-mat.stat-mech/0810.4399}, 
  URL~\url{http://arxiv.org/pdf/0810.4399}.



%\bibitem{my_jps2001}
% I.V. Stasyuk, T.S. Mysakovych,  J. Phys. Studies {\bf 5}, 268 (2001).

%\bibitem{my_cmp2002}
% I.V. Stasyuk, T.S. Mysakovych,  Condens. Matter Phys. {\bf 5}, 473 (2002).



\bibitem{dulepa2}
  I.~V. Stasyuk, I.~R. Dulepa, \emph {Journ. Phys. Studies} {\bf 13}  (2009), in print.


\bibitem{pavl}
  I.~V. Stasyuk, N.~Pavlenko, \emph { J. Phys.: Condens. Matter} {\bf 10}, 7079--7090 (1998).











\end{thebibliography}

%\end{document}
%
%% Support for ukrainian language. Uncomment line above 
%%    if you do not use ukrainian.

\backmatter

\title{Фазов╕ д╕аграми модел╕ Бозе-Хаббарда при ск╕нченн╕й температур╕}

\keywords      {модель Бозе-Хаббарда, фазов╕ переходи}

\author{ ╤.В. Стасюк      }{
  address={     ╤нститут ф╕зики конденсованих систем, 79011 Льв╕в, вул. Свенц╕цького, 1}
}
\author{Т.С. Мисакович }{
%  address={<ЯОЁКЭМЮ ЮДПЕЯЮ ЮБРНПЮ2 Ё ЮБРНПЮ3>}
}

%\author{<ЮБРНП3>}{
%  address={<ЯОЁКЭМЮ ЮДПЕЯЮ ЮБРНПЮ2 Ё ЮБРНПЮ3>}
%  ,altaddress={<ЮДПЕЯЮ ЮБРНПЮ1>} % additional visiting address
%}

\begin{abstract}
 Досл╕джено фазов╕ переходи в модел╕ Бозе-Хаббарда. Одночастинкова функц╕я Гр╕на розрахована в 
 наближенн╕ хаотичних фаз з використанням формал╕зму оператор╕в Хаббарда. 
 Встановлено област╕ ╕снування надплинно╖ фази та фази Мотт╕вського д╕електрика ╕ побудовано 
 фазов╕ д╕аграми в площин╕ $(\mu,t)$. Проанал╕зовано вплив зм╕ни температури на фазов╕ переходи та побудовано фазов╕ д╕аграми в площин╕  $(T,\mu)$. Досл╕джено випадок терм╕чно╖ активац╕╖ перескоку частинок та проанал╕зовано перебудову фазових д╕аграм у цьому випадку.

\end{abstract}

\maketitle

\end{document}